%
%
%
\def\unredoffs{} \def\redoffs{\voffset=-.31truein\hoffset=-.48truein}
\def\speclscape{}
%
%
%
%
%
\newbox\leftpage \newdimen\fullhsize \newdimen\hstitle \newdimen\hsbody
\tolerance=1000\hfuzz=2pt
\catcode`\@=11 
\ifx\hyperdef\UNd@FiNeD\def\hyperdef#1#2#3#4{#4}\def\hyperref#1#2#3#4{#4}\fi
\def\bigans{b }
\def\answ{b }
%
\ifx\answ\bigans\message{(This will come out unreduced.}
\magnification=1200\unredoffs\baselineskip=16pt plus 2pt minus 1pt
\hsbody=\hsize \hstitle=\hsize 
\else\message{(This will be reduced.} \let\l@r=L
\magnification=1000\baselineskip=16pt plus 2pt minus 1pt \vsize=7truein
\redoffs \hstitle=8truein\hsbody=4.75truein\fullhsize=10truein\hsize=\hsbody
\output={\ifnum\pageno=0 
  \shipout\vbox{\speclscape{\hsize\fullhsize\makeheadline}
    \hbox to \fullhsize{\hfill\pagebody\hfill}}\advancepageno
  \else
  \almostshipout{\leftline{\vbox{\pagebody\makefootline}}}\advancepageno
  \fi}
\def\almostshipout#1{\if L\l@r \count1=1 \message{[\the\count0.\the\count1]}
      \global\setbox\leftpage=#1 \global\let\l@r=R
 \else \count1=2
  \shipout\vbox{\speclscape{\hsize\fullhsize\makeheadline}
      \hbox to\fullhsize{\box\leftpage\hfil#1}}  \global\let\l@r=L\fi}
\fi
%
\newcount\yearltd\yearltd=\year\advance\yearltd by -2000

\def\Title#1#2{\nopagenumbers\abstractfont\hsize=\hstitle\rightline{#1}%
\vskip 1in\centerline{\titlefont #2}\abstractfont\vskip .5in\pageno=0}
\def\Date#1{\vfill\leftline{#1}\tenpoint\supereject\global\hsize=\hsbody%
\footline={\hss\tenrm\hyperdef\hypernoname{page}\folio\folio\hss}}%
%

\def\draftmode{\message{ DRAFTMODE }\def\draftdate{{\rm preliminary draft:
\number\month/\number\day/\number\yearltd\ \ \hourmin}}%
\headline={\hfil\draftdate}\writelabels\baselineskip=20pt plus 2pt minus 2pt
 {\count255=\time\divide\count255 by 60 \xdef\hourmin{\number\count255}
  \multiply\count255 by-60\advance\count255 by\time
  \xdef\hourmin{\hourmin:\ifnum\count255<10 0\fi\the\count255}}}
\def\nolabels{\def\wrlabeL##1{}\def\eqlabeL##1{}\def\reflabeL##1{}}
\def\writelabels{\def\wrlabeL##1{\leavevmode\vadjust{\rlap{\smash%
{\line{{\escapechar=` \hfill\rlap{\sevenrm\hskip.03in\string##1}}}}}}}%
\def\eqlabeL##1{{\escapechar-1\rlap{\sevenrm\hskip.05in\string##1}}}%
\def\reflabeL##1{\noexpand\llap{\noexpand\sevenrm\string\string\string##1}}}
\nolabels
%
\global\newcount\secno \global\secno=0
\global\newcount\meqno \global\meqno=1
\def\s@csym{}
\def\newsec#1{\global\advance\secno by1%
{\toks0{#1}\message{(\the\secno. \the\toks0)}}%
\global\subsecno=0\eqnres@t\let\s@csym\secsym\xdef\secn@m{\the\secno}\noindent
{\bf\hyperdef\hypernoname{section}{\the\secno}{\the\secno.} #1}%
\writetoca{{\string\hyperref{}{section}{\the\secno}{\it\the\secno.}} {{\it #1} }}%
\par\nobreak\medskip\nobreak}
\def\eqnres@t{\xdef\secsym{\the\secno.}\global\meqno=1\bigbreak\bigskip}
\def\sequentialequations{\def\eqnres@t{\bigbreak}}\xdef\secsym{}
\global\newcount\subsecno \global\subsecno=0
\def\subsec#1{\global\advance\subsecno by1%
{\toks0{#1}\message{(\s@csym\the\subsecno. \the\toks0)}}%
\ifnum\lastpenalty>9000\else\bigbreak\fi       \global\subsubsecno=0
\noindent{\it\hyperdef\hypernoname{subsection}{\secn@m.\the\subsecno}%
{\secn@m.\the\subsecno.} #1}\writetoca{\string\quad
{\string\hyperref{}{subsection}{\secn@m.\the\subsecno}{\secn@m.\the\subsecno.}}
{#1}}\par\nobreak\medskip\nobreak}
\def\appendix#1#2{\global\meqno=1\global\subsecno=0\xdef\secsym{\hbox{#1.}}%
\bigbreak\bigskip\noindent{\bf Appendix \hyperdef\hypernoname{appendix}{#1}%
{#1.} #2}{\toks0{(#1. #2)}\message{\the\toks0}}%
\xdef\s@csym{#1.}\xdef\secn@m{#1}%
\writetoca{\string\hyperref{}{appendix}{#1}{{\it Appendix} {\it #1.}} {\it #2}}%
\par\nobreak\medskip\nobreak}
%
%
\def\checkm@de#1#2{\ifmmode{\def\f@rst##1{##1}\hyperdef\hypernoname{equation}%
{#1}{#2}}\else\hyperref{}{equation}{#1}{#2}\fi}
\def\eqnn#1{\DefWarn#1\xdef #1{(\noexpand\relax\noexpand\checkm@de%
{\s@csym\the\meqno}{\secsym\the\meqno})}%
\wrlabeL#1\writedef{#1\leftbracket#1}\global\advance\meqno by1}
\def\f@rst#1{\c@t#1a\em@ark}\def\c@t#1#2\em@ark{#1}
\def\eqna#1{\DefWarn#1\wrlabeL{#1$\{\}$}%
\xdef #1##1{(\noexpand\relax\noexpand\checkm@de%
{\s@csym\the\meqno\noexpand\f@rst{##1}}{\hbox{$\secsym\the\meqno##1$}})}
\writedef{#1\numbersign1\leftbracket#1{\numbersign1}}\global\advance\meqno by1}
\def\eqn#1#2{\DefWarn#1%
\xdef #1{(\noexpand\hyperref{}{equation}{\s@csym\the\meqno}%
{\secsym\the\meqno})}$$#2\eqno(\hyperdef\hypernoname{equation}%
{\s@csym\the\meqno}{\secsym\the\meqno})\eqlabeL#1$$%
\writedef{#1\leftbracket#1}\global\advance\meqno by1}
\def\xeqn{\expandafter\xe@n}\def\xe@n(#1){#1}
\def\xeqna#1{\expandafter\xe@n#1}
\def\eqns#1{(\e@ns #1{\hbox{}})}
\def\e@ns#1{\ifx\UNd@FiNeD#1\message{eqnlabel \string#1 is undefined.}%
\xdef#1{(?.?)}\fi{\let\hyperref=\relax\xdef\next{#1}}%
\ifx\next\em@rk\def\next{}\else%
\ifx\next#1\xeqn#1\else\def\n@xt{#1}\ifx\n@xt\next#1\else\xeqna#1\fi
\fi\let\next=\e@ns\fi\next}

\def\DefWarn#1{\ifx\UNd@FiNeD#1\else
\immediate\write16{*** WARNING: the label \string#1 is already defined ***}\fi}
%
\newskip\footskip\footskip14pt plus 1pt minus 1pt 
\def\footnotefont{\ninepoint}\def\f@t#1{\footnotefont #1\@foot}
\def\f@@t{\baselineskip\footskip\bgroup\footnotefont\aftergroup\@foot\let\next}
\setbox\strutbox=\hbox{\vrule height9.5pt depth4.5pt width0pt}
\global\newcount\ftno \global\ftno=0
\def\foot{\global\advance\ftno by1\def\foot@rg{\hyperref{}{footnote}%
{\the\ftno}{\the\ftno}\xdef\foot@rg{\noexpand\hyperdef\noexpand\hypernoname%
{footnote}{\the\ftno}{\the\ftno}}}\footnote{$^{\foot@rg}$}}
%
\newwrite\ftfile
\def\footend{\def\foot{\global\advance\ftno by1\chardef\wfile=\ftfile
\hyperref{}{footnote}{\the\ftno}{$^{\the\ftno}$}%
\ifnum\ftno=1\immediate\openout\ftfile=\jobname.fts\fi%
\immediate\write\ftfile{\noexpand\smallskip%
\noexpand\item{\noexpand\hyperdef\noexpand\hypernoname{footnote}
{\the\ftno}{f\the\ftno}:\ }\pctsign}\findarg}%
\def\footatend{\vfill\eject\immediate\closeout\ftfile{\parindent=20pt
\centerline{\bf Footnotes}\nobreak\bigskip\input \jobname.fts }}}
\def\footatend{}
%
%
\global\newcount\refno \global\refno=1
\newwrite\rfile
\def\ref{[\hyperref{}{reference}{\the\refno}{\the\refno}]\nref}
\def\nref#1{\DefWarn#1%
\xdef#1{[\noexpand\hyperref{}{reference}{\the\refno}{\the\refno}]}%
\writedef{#1\leftbracket#1}%
\ifnum\refno=1\immediate\openout\rfile=\jobname.refs\fi
\chardef\wfile=\rfile\immediate\write\rfile{\noexpand\item{[\noexpand\hyperdef%
\noexpand\hypernoname{reference}{\the\refno}{\the\refno}]\ }%
\reflabeL{#1\hskip.31in}\pctsign}\global\advance\refno by1\findarg}
\def\findarg#1#{\begingroup\obeylines\newlinechar=`\^^M\pass@rg}
{\obeylines\gdef\pass@rg#1{\writ@line\relax #1^^M\hbox{}^^M}%
\gdef\writ@line#1^^M{\expandafter\toks0\expandafter{\striprel@x #1}%
\edef\next{\the\toks0}\ifx\next\em@rk\let\next=\endgroup\else\ifx\next\empty%
\else\immediate\write\wfile{\the\toks0}\fi\let\next=\writ@line\fi\next\relax}}
\def\striprel@x#1{} \def\em@rk{\hbox{}}
\def\lref{\begingroup\obeylines\lr@f}
\def\lr@f#1#2{\DefWarn#1\gdef#1{\let#1=\UNd@FiNeD\ref#1{#2}}\endgroup\unskip}

\def\addref#1{\immediate\write\rfile{\noexpand\item{}#1}} 
\def\listrefs{\footatend\vfill\supereject\immediate\closeout\rfile\writestoppt
\baselineskip=\footskip\centerline{{\bf References}}\bigskip{\parindent=20pt%
\frenchspacing\escapechar=` \input \jobname.refs\vfill\eject}\nonfrenchspacing}
\def\startrefs#1{\immediate\openout\rfile=\jobname.refs\refno=#1}
\def\xref{\expandafter\xr@f}\def\xr@f[#1]{#1}
\def\refs#1{\count255=1[\r@fs #1{\hbox{}}]}
\def\r@fs#1{\ifx\UNd@FiNeD#1\message{reflabel \string#1 is undefined.}%
\nref#1{need to supply reference \string#1.}\fi%
\vphantom{\hphantom{#1}}{\let\hyperref=\relax\xdef\next{#1}}%
\ifx\next\em@rk\def\next{}%
\else\ifx\next#1\ifodd\count255\relax\xref#1\count255=0\fi%
\else#1\count255=1\fi\let\next=\r@fs\fi\next}
%

%
\newwrite\ffile\global\newcount\figno \global\figno=1
\def\fig{fig.~\hyperref{}{figure}{\the\figno}{\the\figno}\nfig}
\def\nfig#1{\DefWarn#1%
\xdef#1{fig.~\noexpand\hyperref{}{figure}{\the\figno}{\the\figno}}%
\writedef{#1\leftbracket fig.\noexpand~\xfig#1}%
\ifnum\figno=1\immediate\openout\ffile=\jobname.figs\fi\chardef\wfile=\ffile%
{\let\hyperref=\relax
\immediate\write\ffile{\noexpand\medskip\noexpand\item{Fig.\ %
\noexpand\hyperdef\noexpand\hypernoname{figure}{\the\figno}{\the\figno}. }
\reflabeL{#1\hskip.55in}\pctsign}}\global\advance\figno by1\findarg}
\def\listfigs{\vfill\eject\immediate\closeout\ffile{\parindent40pt
\baselineskip14pt\centerline{{\bf Figure Captions}}\nobreak\medskip
\escapechar=` \input \jobname.figs\vfill\eject}}
\def\xfig{\expandafter\xf@g}\def\xf@g fig.\penalty\@M\ {}
\def\figs#1{figs.~\f@gs #1{\hbox{}}}
\def\f@gs#1{{\let\hyperref=\relax\xdef\next{#1}}\ifx\next\em@rk\def\next{}\else
\ifx\next#1\xfig #1\else#1\fi\let\next=\f@gs\fi\next}
\def\figin{\epsfcheck\figin}\def\figins{\epsfcheck\figins}
\def\epsfcheck{\ifx\epsfbox\UNd@FiNeD
\message{(NO epsf.tex, FIGURES WILL BE IGNORED)}
\gdef\figin##1{\vskip2in}\gdef\figins##1{\hskip.5in}
\else\message{(FIGURES WILL BE INCLUDED)}%
\gdef\figin##1{##1}\gdef\figins##1{##1}\fi}
\def\DefWarn#1{}
\def\figinsert{\goodbreak\midinsert}
\def\ifig#1#2#3{\DefWarn#1\xdef#1{Fig.~\noexpand\hyperref{}{figure}%
{\the\figno}{\the\figno}}\writedef{#1\leftbracket fig.\noexpand~\xfig#1}%
\figinsert\figin{\centerline{#3}}\medskip\centerline{\vbox{\baselineskip12pt
\advance\hsize by -1truein\noindent\wrlabeL{#1=#1}\footnotefont%
{\bf Fig.~\hyperdef\hypernoname{figure}{\the\figno}{\the\figno}:} #2}}
\bigskip\endinsert\global\advance\figno by1}
\newwrite\lfile
{\escapechar-1\xdef\pctsign{\string\%}\xdef\leftbracket{\string\{}
\xdef\rightbracket{\string\}}\xdef\numbersign{\string\#}}
\def\writedefs{\immediate\openout\lfile=\jobname.defs \def\writedef##1{%
{\let\hyperref=\relax\let\hyperdef=\relax\let\hypernoname=\relax
 \immediate\write\lfile{\string\def\string##1\rightbracket}}}}%
\def\writestop{\def\writestoppt{\immediate\write\lfile{\string\pageno
 \the\pageno\string\startrefs\leftbracket\the\refno\rightbracket
 \string\def\string\secsym\leftbracket\secsym\rightbracket
 \string\secno\the\secno\string\meqno\the\meqno}\immediate\closeout\lfile}}
\def\writestoppt{}\def\writedef#1{}
\def\seclab#1{\DefWarn#1%
\xdef #1{\noexpand\hyperref{}{section}{\the\secno}{\the\secno}}%
\writedef{#1\leftbracket#1}\wrlabeL{#1=#1}}
\def\subseclab#1{\DefWarn#1%
\xdef #1{\noexpand\hyperref{}{subsection}{\secn@m.\the\subsecno}%
{\secn@m.\the\subsecno}}\writedef{#1\leftbracket#1}\wrlabeL{#1=#1}}
\def\applab#1{\DefWarn#1%
\xdef #1{\noexpand\hyperref{}{appendix}{\secn@m}{\secn@m}}%
\writedef{#1\leftbracket#1}\wrlabeL{#1=#1}}
\newwrite\tfile \def\writetoca#1{}
\def\leaderfill{\leaders\hbox to 1em{\hss.\hss}\hfill}
\def\writetoc{\immediate\openout\tfile=\jobname.toc
   \def\writetoca##1{{\edef\next{\write\tfile{\noindent ##1
   \string\leaderfill {\string\hyperref{}{page}{\noexpand\number\pageno}%
                       {\noexpand\number\pageno}} \par}}\next}}}
\newread\ch@ckfile
\def\listtoc{\immediate\closeout\tfile\immediate\openin\ch@ckfile=\jobname.toc
\ifeof\ch@ckfile\message{no file \jobname.toc, no table of contents this pass}%
\else\closein\ch@ckfile\centerline{\bf Contents}\nobreak\medskip%
{\baselineskip=18.5pt  \footnotefont
\parskip=2pt\catcode`\@=12\input\jobname.toc
\catcode`\@=12\bigbreak\bigskip}\fi}
\catcode`\@=12 
%
\edef\tfontsize{\ifx\answ\bigans scaled\magstep3\else scaled\magstep4\fi}
\font\titlerm=cmr10 \tfontsize \font\titlerms=cmr7 \tfontsize
\font\titlermss=cmr5 \tfontsize \font\titlei=cmmi10 \tfontsize
\font\titleis=cmmi7 \tfontsize \font\titleiss=cmmi5 \tfontsize
\font\titlesy=cmsy10 \tfontsize \font\titlesys=cmsy7 \tfontsize
\font\titlesyss=cmsy5 \tfontsize \font\titleit=cmti10 \tfontsize
\skewchar\titlei='177 \skewchar\titleis='177 \skewchar\titleiss='177
\skewchar\titlesy='60 \skewchar\titlesys='60 \skewchar\titlesyss='60
\def\titlefont{\def\rm{\fam0\titlerm}
\textfont0=\titlerm \scriptfont0=\titlerms \scriptscriptfont0=\titlermss
\textfont1=\titlei \scriptfont1=\titleis \scriptscriptfont1=\titleiss
\textfont2=\titlesy \scriptfont2=\titlesys \scriptscriptfont2=\titlesyss
\textfont\itfam=\titleit \def\it{\fam\itfam\titleit}\rm}
 \ifx\answ\bigans\else scaled\magstep1\fi
\ifx\answ\bigans\def\abstractfont{\tenpoint}\else
\font\absit=cmti10 scaled \magstep1
\font\abssl=cmsl10 scaled \magstep1
\font\absrm=cmr10 scaled\magstep1 \font\absrms=cmr7 scaled\magstep1
\font\absrmss=cmr5 scaled\magstep1 \font\absi=cmmi10 scaled\magstep1
\font\absis=cmmi7 scaled\magstep1 \font\absiss=cmmi5 scaled\magstep1
\font\abssy=cmsy10 scaled\magstep1 \font\abssys=cmsy7 scaled\magstep1
\font\abssyss=cmsy5 scaled\magstep1 \font\absbf=cmbx10 scaled\magstep1
\skewchar\absi='177 \skewchar\absis='177 \skewchar\absiss='177
\skewchar\abssy='60 \skewchar\abssys='60 \skewchar\abssyss='60
\def\abstractfont{\def\rm{\fam0\absrm}
\textfont0=\absrm \scriptfont0=\absrms \scriptscriptfont0=\absrmss
\textfont1=\absi \scriptfont1=\absis \scriptscriptfont1=\absiss
\textfont2=\abssy \scriptfont2=\abssys \scriptscriptfont2=\abssyss
\textfont\itfam=\absit \def\it{\fam\itfam\absit}\def\footnotefont{\tenpoint}%
\textfont\slfam=\abssl \def\sl{\fam\slfam\abssl}%
\textfont\bffam=\absbf \def\bf{\fam\bffam\absbf}\rm}\fi
\def\tenpoint{\def\rm{\fam0\tenrm}
\textfont0=\tenrm \scriptfont0=\sevenrm \scriptscriptfont0=\fiverm
\textfont1=\teni  \scriptfont1=\seveni  \scriptscriptfont1=\fivei
\textfont2=\tensy \scriptfont2=\sevensy \scriptscriptfont2=\fivesy
\textfont\itfam=\tenit \def\it{\fam\itfam\tenit}\def\footnotefont{\ninepoint}%
\textfont\bffam=\tenbf \def\bf{\fam\bffam\tenbf}\def\sl{\fam\slfam\tensl}\rm}
\font\ninerm=cmr9 \font\sixrm=cmr6 \font\ninei=cmmi9 \font\sixi=cmmi6
\font\ninesy=cmsy9 \font\sixsy=cmsy6 \font\ninebf=cmbx9
\font\nineit=cmti9 \font\ninesl=cmsl9 \skewchar\ninei='177
\skewchar\sixi='177 \skewchar\ninesy='60 \skewchar\sixsy='60
\def\ninepoint{\def\rm{\fam0\ninerm}
\textfont0=\ninerm \scriptfont0=\sixrm \scriptscriptfont0=\fiverm
\textfont1=\ninei \scriptfont1=\sixi \scriptscriptfont1=\fivei
\textfont2=\ninesy \scriptfont2=\sixsy \scriptscriptfont2=\fivesy
\textfont\itfam=\ninei \def\it{\fam\itfam\nineit}\def\sl{\fam\slfam\ninesl}%
\textfont\bffam=\ninebf \def\bf{\fam\bffam\ninebf}\rm}
%
%
\def\noblackbox{\overfullrule=0pt}
\hyphenation{anom-aly anom-alies coun-ter-term coun-ter-terms}
\def\inv{^{\raise.15ex\hbox{${\scriptscriptstyle -}$}\kern-.05em 1}}

\def\Dsl{\,\raise.15ex\hbox{/}\mkern-13.5mu D} 
\def\dsl{\raise.15ex\hbox{/}\kern-.57em\partial}

\def\lspace{\ifx\answ\bigans{}\else\qquad\fi}
\def\lbspace{\ifx\answ\bigans{}\else\hskip-.2in\fi} 
\def\boxeqn#1{\vcenter{\vbox{\hrule\hbox{\vrule\kern3pt\vbox{\kern3pt
	\hbox{${\displaystyle #1}$}\kern3pt}\kern3pt\vrule}\hrule}}}
\def\mbox#1#2{\vcenter{\hrule \hbox{\vrule height#2in
		\kern#1in \vrule} \hrule}}  
%

\def\darr#1{\raise1.5ex\hbox{$\leftrightarrow$}\mkern-16.5mu #1}

\def\roughly#1{\raise.3ex\hbox{$#1$\kern-.75em\lower1ex\hbox{$\sim$}}}

\global\newcount\subsubsecno \global\subsubsecno=0
\def\subsubsec#1{\global\advance\subsubsecno by1%
{\toks0{#1}\message{(\the\secno\the\subsecno\the\subsubsecno. \the\toks0)}}%
\ifnum\lastpenalty>9000\else\bigbreak\fi
\noindent{\it\hyperdef\hypernoname{subsubsection}{\the\secno.\the\subsecno\the\subsubsecno}%
{\the\secno.\the\subsecno.\the\subsubsecno.} #1}
\par\nobreak\medskip\nobreak}
\def\boxit#1{\vbox{\hrule\hbox{\vrule\kern8pt
\vbox{\hbox{\kern8pt}\hbox{\vbox{#1}}\hbox{\kern8pt}}
\kern8pt\vrule}\hrule}}
\def\mathboxit#1{\vbox{\hrule\hbox{\vrule\kern8pt\vbox{\kern8pt
\hbox{$\displaystyle #1$}\kern8pt}\kern8pt\vrule}\hrule}}
\def\slashchar#1{\setbox0=\hbox{$#1$}           
   \dimen0=\wd0                                 
   \setbox1=\hbox{/} \dimen1=\wd1               
   \ifdim\dimen0>\dimen1                        
      \rlap{\hbox to \dimen0{\hfil/\hfil}}      
      #1                                        
   \else                                        
      \rlap{\hbox to \dimen1{\hfil$#1$\hfil}}   
      /                                         
   \fi}
\def\sqr#1#2{{\vcenter{\vbox{\hrule height.#2pt
         \hbox{\vrule width.#2pt height#1pt \kern#1pt
            \vrule width.#2pt}
         \hrule height.#2pt}}}}


\input amssym.def
\input amssym.tex
\noblackbox
\baselineskip=14.5pt
\def\crr{\noalign{\vskip5pt}}

\def\comment#1{{}}

\def\ap{\alpha'}

\def\cf{{\it cf.\ }}
\def\ie{{\it i.e.\ }}

\def\al{\alpha}

\newif\ifnref

\nreffalse

\input epsf

\def\figin{\epsfcheck\figin}\def\figins{\epsfcheck\figins}
\def\epsfcheck{\ifx\epsfbox\UnDeFiNeD
\message{(NO epsf.tex, FIGURES WILL BE IGNORED)}
\gdef\figin##1{\vskip2in}\gdef\figins##1{\hskip.5in}
\else\message{(FIGURES WILL BE INCLUDED)}%
\gdef\figin##1{##1}\gdef\figins##1{##1}\fi}
\def\DefWarn#1{}
\def\figinsert{\goodbreak\midinsert}  
\def\ifig#1#2#3{\DefWarn#1\xdef#1{Fig.~\the\figno}
\writedef{#1\leftbracket fig.\noexpand~\the\figno}%
\figinsert\figin{\centerline{#3}}\medskip\centerline{\vbox{\baselineskip12pt
\advance\hsize by -1truein\noindent\footnotefont\centerline{{\bf
Fig.~\the\figno}\ \sl #2}}}
\bigskip\endinsert\global\advance\figno by1}

\def\iifig#1#2#3#4{\DefWarn#1\xdef#1{Fig.~\the\figno}
\writedef{#1\leftbracket fig.\noexpand~\the\figno}%
\figinsert\figin{\centerline{#4}}\medskip\centerline{\vbox{\baselineskip12pt
\advance\hsize by -1truein\noindent\footnotefont\centerline{{\bf
Fig.~\the\figno}\ \ \sl #2}}}\smallskip\centerline{\vbox{\baselineskip12pt
\advance\hsize by -1truein\noindent\footnotefont\centerline{\ \ \ \sl #3}}}
\bigskip\endinsert\global\advance\figno by1}


\def\h {{1\over 2}}

\def\ov {\overline}
\def\o {\over}
\def\fc#1#2{{#1 \o #2}}

\def\IC{{\bf C}}


\def\br{\hfill\break}

\def\det {{\rm det}}
\def\mod {{\rm mod}}
\def\lf {\left}
\def\ri {\right}

\def\re {{\rm Re}}
\def\im {{\rm Im}}
\def\p {\partial}

 \def\Oc {{\cal O}}

\def\ceiling#1{\lceil#1\rceil}
\def\floor#1{\lfloor#1\rfloor}


\def\sv{{\rm sv}}


\lref\StiebergerWEA{
  S.~Stieberger,
``Closed superstring amplitudes, single-valued multiple zeta values and the Deligne associator,''
J.\ Phys.\ A {\bf 47}, 155401 (2014).
[arXiv:1310.3259 [hep-th]].
}

\lref\StiebergerHBA{
  S.~Stieberger and T.R.~Taylor,
 ``Closed String Amplitudes as Single-Valued Open String Amplitudes,''
Nucl.\ Phys.\ B {\bf 881}, 269 (2014).
[arXiv:1401.1218 [hep-th]].
}

\lref\BernQJ{
  Z.~Bern, J.J.M.~Carrasco and H.~Johansson,
``New Relations for Gauge-Theory Amplitudes,''
Phys.\ Rev.\ D {\bf 78}, 085011 (2008).
[arXiv:0805.3993 [hep-ph]].
}

\lref\StiebergerHQ{
  S.~Stieberger,
``Open \& Closed vs. Pure Open String Disk Amplitudes,''
[arXiv:0907.2211 [hep-th]].
}

\lref\stnew{
  S.~Stieberger and T.R.~Taylor, in preparation.}

\lref\KawaiXQ{
  H.~Kawai, D.C.~Lewellen and S.H.H.~Tye,
``A Relation Between Tree Amplitudes Of Closed And Open Strings,''
  Nucl.\ Phys.\  B {\bf 269}, 1 (1986).
}

\lref\StiebergerTE{
  S.~Stieberger and T.R.~Taylor,
``Multi-Gluon Scattering in Open Superstring Theory,''
Phys.\ Rev.\ D {\bf 74}, 126007 (2006).
[hep-th/0609175].
}

\lref\MafraNVi{
  C.R.~Mafra, O.~Schlotterer and S.~Stieberger,
``Complete N-Point Superstring Disk Amplitude I. Pure Spinor Computation,''
Nucl.\ Phys.\ B {\bf 873}, 419 (2013).
[arXiv:1106.2645 [hep-th]];
``Complete N-Point Superstring Disk Amplitude II. Amplitude and Hypergeometric Function Structure,''
Nucl.\ Phys.\ B {\bf 873}, 461 (2013).
[arXiv:1106.2646 [hep-th]].
}

\lref\SiegelSK{
  W.~Siegel,
  ``Hidden gravity in open string field theory,''
Phys.\ Rev.\ D {\bf 49}, 4144 (1994).
[hep-th/9312117].
}
\lref\ChiodaroliXIA{
  M.~Chiodaroli,  M.~Gunaydin, H.~Johansson and R.~Roiban,
  ``Scattering amplitudes in N=2 Maxwell-Einstein and Yang-Mills/Einstein supergravity,''
[arXiv:1408.0764 [hep-th]].
}

\lref\ChenCT{
  Y.X.~Chen, Y.J.~Du and B.~Feng,
  ``On tree amplitudes with gluons coupled to gravitons,''
JHEP {\bf 1101}, 081 (2011).
[arXiv:1011.1953 [hep-th]].
}

\lref\BrittoFQ{
  R.~Britto, F.~Cachazo, B.~Feng and E.~Witten,
  ``Direct proof of tree-level recursion relation in Yang-Mills theory,''
Phys.\ Rev.\ Lett.\  {\bf 94}, 181602 (2005).
[hep-th/0501052].
}

\lref\notation{M.L.~Mangano and S.J.~Parke,
``Multiparton amplitudes in gauge theories,''
Phys. Rept.  {\bf 200}, 301 (1991).
[hep-th/0509223];\br
L.J.~Dixon,
  ``Calculating scattering amplitudes efficiently,''
in Boulder 1995, QCD and beyond 539-582.
[hep-ph/9601359].}

\lref\sakh{A.D.\ Sakharov, ``Vacuum Quantum Fluctuations in Curved Space
and the Theory of Gravitation,''   Dokl.\ Akad.\ Nauk SSSR {\bf 177}, 70 (1967) [Gen.\ Rel.\ Grav.\ {\bf 32}, 365 (2000)].}

\lref\Veneziano{D.~Amati and G.~Veneziano, ``Metric From Matter,''
Phys.\ Lett.\ B {\bf 105}, 358 (1981);\br
S.L.~Adler,
  ``Einstein Gravity as a Symmetry Breaking Effect in Quantum Field Theory,''
Rev.\ Mod.\ Phys.\  {\bf 54}, 729 (1982), [Erratum-ibid.\  {\bf 55}, 837 (1983)].}

\lref\KosteleckyPX{
  V.A.~Kostelecky, O.~Lechtenfeld and S.~Samuel,
``Covariant String Amplitudes On Exotic Topologies To One Loop,''
Nucl.\ Phys.\ B {\bf 298}, 133 (1988).
}

\lref\StiebergerTE{
  S.~Stieberger and T.R.~Taylor,
``Multi-Gluon Scattering in Open Superstring Theory,''
Phys.\ Rev.\ D {\bf 74}, 126007 (2006).
[hep-th/0609175].
}

\lref\CachazoFWA{
  F.~Cachazo and A.~Strominger,
``Evidence for a New Soft Graviton Theorem,''
[arXiv:1404.4091 [hep-th]].
}

\lref\CasaliXPA{
  E.~Casali,
``Soft sub-leading divergences in Yang-Mills amplitudes,''
JHEP {\bf 1408}, 077 (2014).
[arXiv:1404.5551 [hep-th]].
}
\lref\BernBX{
  Z.~Bern, A.~De Freitas and H.L.~Wong,
``On the coupling of gravitons to matter,''
Phys.\ Rev.\ Lett.\  {\bf 84}, 3531 (2000).
[hep-th/9912033].
}
\lref\WeinbergKQ{
  S.~Weinberg and E.~Witten,
  ``Limits on Massless Particles,''
Phys.\ Lett.\ B {\bf 96}, 59 (1980).
}

\Title{\vbox{\rightline{MPP--2014--343}
}}
{\vbox{\centerline{Graviton as a Pair of Collinear Gauge Bosons}}}
\medskip
\centerline{Stephan Stieberger$^a$ and Tomasz R. Taylor$^b$}
\bigskip
\centerline{\it $^a$ Max--Planck--Institut f\"ur Physik}
\centerline{\it Werner--Heisenberg--Institut, 80805 M\"unchen, Germany}
\medskip
\centerline{\it  $^b$ Department of Physics}
\centerline{\it  Northeastern University, Boston, MA 02115, USA}

\vskip15pt

\medskip
\bigskip\bigskip\bigskip
\centerline{\bf Abstract}
\vskip .2in
\noindent

\noindent
We show that the mixed gravitational/gauge superstring amplitudes describing decays of massless closed strings -- gravitons or dilatons -- into a number of gauge bosons, can be written at the tree (disk) level as  linear combinations of pure open string amplitudes in which the graviton (or dilaton) is replaced by a pair of collinear gauge bosons. Each of the constituent gauge bosons carry exactly one half of the original closed string momentum, while their $\pm 1$ helicities add up to $\pm 2$ for the graviton or to 0 for the dilaton.

\Date{}
\noindent
\goodbreak
\break

Quantization of gravitational waves yields gravitons: massless spin 2 particles with two polarized degrees of freedom (helicity ${+}2\equiv ++$ and ${-}2\equiv --$) in four dimensions. While the existence of gravitational waves is well established, the detection of individual gravitons may be impossible due to extremely low cross sections. Nevertheless, theoretical understanding of  gravitons and their interactions is a  prerequisite for constructing a viable theory of quantum gravity.

Superstring theory offers an interesting insight into gravitons. In this framework, they appear as zero modes of {\it closed\/} strings. On the other hand, it is known that zero modes of {\it open\/} strings give rise to spin 1 gauge bosons. With the closed string seen as a loop of two  open strings connected at both ends, graviton appears to be a ``bound state'' of two vector bosons. This  is also suggested by the form of graviton vertex operator: in type II superstring theory, it is a product of two spin 1 vertex operators (from the left- and right-moving sectors of world-sheet excitations). Helicity $++$ appears as a superposition of two helicity $+$  states while helicity $--$ comes as a superposition of two helicity $-$ states. In addition, the products $+-$ and $-+$ create two degrees of freedom of the scalar (complex) superstring dilaton.

In 1985, Kawai, Lewellen and Tye (KLT) \KawaiXQ\ derived a formula which expresses any closed string tree amplitude in terms of a sum of the products of appropriate open string tree amplitudes. At the level of zero modes,  KLT relations allow expressing the graviton and dilaton amplitudes in terms of products of gauge boson amplitudes. The existence of such relations means that, at least in the leading order of perturbation theory, the content of Einstein's gravity is encoded in Yang--Mills (YM)  theory. The quadratic form of KLT relations is perfectly consistent with the heuristic picture of a closed string as a loop of two open strings. In fact, string field theory suggests a similar description \SiegelSK. This does not help, however, in answering the question whether the graviton can be considered as a pair of gauge bosons beyond the world--sheet, as an actual bound state in physical space-time. One alternative description has been developed in
\refs{\StiebergerWEA,\StiebergerHBA}, by constructing closed superstring amplitudes through the  ``single--valued'' projection of open superstring amplitudes. This projection yields linear relations between the functions encompassing effects of massive closed and open superstring excitations,
to all orders in the inverse string tension $\ap$.
They reveal a deeper connection between gauge and gravity string amplitudes than what is implied by the KLT relations, but they do not provide new insight into their $\ap\to 0$ field theory limit.

In this Letter, we present a {\it linear\/} relation between the amplitude for the decay of one massless closed string state, {\it i.e}.\ a graviton or a dilaton, into an arbitrary number $N{-}2$ of gauge bosons and a sum of purely open string amplitudes involving $N$ gauge bosons. The sum involves so-called partial amplitudes associated to particular gauge group factors. The original closed string state is replaced by two vector bosons, each of them carrying exactly one half of its momentum, and its helicity is split in the same way as in string vertex operators. In the forthcoming publication \stnew, we will show that in all open and closed string amplitudes, gravitons and dilatons can be  replaced by pairs of such collinear vectors bosons.

Although our derivation utilizes full-fledged Type II superstring theory, it is instructive to discuss the field theory limit ({\it i.e}.\ the zero slope $\alpha'=0$ limit of Regge trajectories) of mixed gravitational/gauge interactions. This limit is described by Einstein-Yang-Mills (EYM) theory  coupled to the dilaton\foot{For early work on EYM scattering amplitudes, see Ref. \BernBX; for more recent work, see  \ChiodaroliXIA.}. All tree level amplitudes can be constructed by using the recursion relations derived by Britto, Cachazo, Feng and Witten (BCFW) \BrittoFQ, with the basic building blocks provided by the following three-point amplitudes:
\eqn\three{\eqalign{A(1^{--},2^{--},3^{++})={\langle 12\rangle^6\over\langle 23\rangle^2\langle 31\rangle^2}\ ,& \qquad A(1^{--},2^{+-},3^{-+})={\langle 12\rangle^2\langle 13\rangle^2\over\langle 23\rangle^2}
\cr\crr A(1^{+},2^{-},3^{--})={\langle 23\rangle^4\over\langle 12\rangle^2}\ ,\qquad & \qquad ~~~~A(1^{-},2^{-},3^{+-})=\langle 12\rangle^2\ ,
}}
where we used superscripts to label helicity states, with $+-$ and $-+$ assigned to the dilaton and its complex conjugate, respectively. We are using standard notation of the helicity formalism, see \notation. The mass dimension $(-1)$ gravitational coupling $\sqrt \kappa$ is implied by the above expressions. In addition, three gauge bosons interact with the well known Yang-Mills amplitude
\eqn\ym{A(1^{-},2^{-},3^{+})={\langle 12\rangle^3\over\langle 23\rangle\langle 31\rangle}\ , }
where we omitted the (dimensionless) gauge coupling constant.

A good example of an amplitude involving both gravitational and gauge couplings is the amplitude for  the graviton decay into three gauge bosons. In this case,
\eqn\example{A(1^{+},2^{+},3^{-};q^{--})={\langle 3q\rangle^4\over\langle 12\rangle\langle 23\rangle\langle 31\rangle}\ , }
which can be obtained either by using BCFW recursion relations or by a straightforward Feynman diagram
 calculation. In this Letter, we focus on the amplitudes similar to \example, describing gravitons and dilatons decaying into an arbitrary number of gauge bosons. In string theory, these are disk amplitudes with one closed string vertex insertion on the world-sheet and a number of open strings attached at the boundary.

In order to compute the amplitudes, it is convenient to use the ``doubling trick,''
to convert disk correlators to the standard holomorphic ones by extending the fields to the entire complex plane \KosteleckyPX. Furthermore, the integration over  positions of  world--sheet symmetric closed string states (such as graviton or dilaton) can be extended from the half--plane covering the disk to the full complex plane. Open string vertices representing $N{-}2\,$ gauge bosons with momenta $p_i,~i=1,\dots,N{-}2$ (in an arbitrary helicity configuration) are inserted on the real axis at $x_i$,  while a single closed string vertex operator, which represents the graviton or dilaton with momentum $q$, is inserted at complex $z$. All momenta are restricted to four  dimensions,
 with  $p_i^2=q^2=0$ (although the following derivation is independent on the space--time dimension and can be adjusted to massive states). The amplitudes involve integrals of the form
\eqn\GENERIC{\eqalign{
F_N&=V_{\rm CKG}^{-1}\ \delta^{(4)}\Big(\sum_{i=1}^{N-2}p_i+q\Big)\int\prod_{i=1}^{N-2} dx_i
\prod_{1\leq r<s\leq N-2}|x_r-x_s|^{2\alpha'p_rp_s}\ (x_r-x_s)^{n_{rs}}\cr
&\times \int_{\IC} d^2z\ (z-\ov z)^n \ \prod_{i=1}^{N-2}\ |x_i-z|^{2\al'p_iq}\ (x_i-z)^{n_i}\ (x_i-\ov z)^{\bar n_i}\ ,}}
where we included the momentum-conserving delta function and divided by the volume $V_{\rm CKG}$ of the conformal Killing group.
The powers $n_{rs},~n_i,~\ov n_i,~ n$ are some integer numbers. To be specific, we focus on the amplitude associated to
one particular Chan-Paton factor (partial amplitude), ${\rm Tr}(T^1T^2\dots T^{N-2})$, with the integral over ordered  $x_1<x_2<\dots<x_{N{-}2}$.

The techniques for evaluating generic disk integrals involving both open and closed strings have been
developed in \StiebergerHQ.
For the concrete case \GENERIC, we write the complex integral as an integral
over holomorphic and anti--holomorphic coordinates, by following the method proposed in \KawaiXQ.
After writing
$z=z_{1}+i z_{2}$, the integrand becomes an analytic function of $z_{2}$ with $2(N{-}2)$ branch points
at $\pm i(x_i-z_1)$.
We then deform the $z_{2}$--integral along the real axis $\im(z_{2})=0$ to the
pure imaginary axis $\re(z_{2})=0$.
In this way, the variables
\eqn\real{
\xi=z_{1}+i\ z_{2}\equiv z\ \ \ ,\ \ \ \eta=z_{1}-i\ z_{2}\equiv \ov z}
become real.
After changing the integration variables $(z_1,z_2)\to (\xi,\eta)$ (with the Jacobian  $\det\fc{\p(z_{1},z_{2})}{\p(\xi,\eta)}=\fc{i}{2}$), Eq. \GENERIC\ becomes an integral over $N$ real positions $x_i,\xi,\eta$
\eqn\AMPLITUDE{\eqalign{
F_N&=V_{\rm CKG}^{-1}\ \delta^{(4)}\Big(\sum_{i=1}^{N}k_i\Big)\int\prod_{i=1}^{N-2} dx_i
\int_{-\infty}^\infty d\xi\int_{-\infty}^\infty d\eta\!
\prod_{1\leq r<s\leq N-2}|x_r-x_s|^{2\alpha'k_rk_s}\ (x_r-x_s)^{n_{rs}}\cr
&\times\fc{i}{2}\  (\xi-\eta)^n\prod_{i=1}^{N-2} \Pi(x_i,\xi,\eta)\  |x_i-\xi|^{2\alpha'k_ik_{N{-}1}}\ |x_i-\eta|^{2\alpha'k_ik_N}(x_i-\xi)^{n_i}(x_i-\eta)^{\bar n_i}\ ,}}
with the open string momenta $k_r=p_r,\ r=1,\ldots,N{-}2$ and the closed string momentum split in half:
\eqn\split{
k_{N{-}1}=k_N={1\over 2}\, q\ .}
Eq. \AMPLITUDE\ resembles a generic  open string integral involving $N$ open strings with external
momenta $k_i$ supplemented by the  extra phase factors
\eqn\PHASE{
\Pi(x_i,\xi,\eta)=e^{2\pi i\al'k_ik_N \,\theta[-(x_i-\xi)(x_i-\eta)]}\ ,}
where $\theta$ denotes the Heaviside step function.
These monodromy factors \PHASE\ account for the
correct branch of the integrand,
making the integral well defined.
 Note that the phases, which are independent on the integers
$n_{rs},n_i,\ov n_i, n$ do not depend on the particular values of  integration variables, but only on the ordering of $\xi$ and $\eta$ with respect to the original $N{-}2$ vertex positions.
In this way, the original integral becomes a weighted (by phase factors) sum of integrals, each of them having the same form as the integrals appearing in $N$-point (partial) open string amplitudes, with the vertices inserted at  $x_l, ~l=1,\dots,N$, where we identified $x_{N{-}1}\equiv\xi$  and $x_N\equiv\eta$.
Note that the order of the original $N{-}2$ positions remains unchanged. Since the graviton as well as dilaton vertices factorize into two gauge bosons inserted at $z=\xi=x_{N{-}1}$ and  $\ov z=\eta=x_N $, we conclude that
the amplitude $A(1,2,\dots,N{-}2;q)$ describing graviton (or dilaton) decays into  $N{-}2$ gauge bosons can be written as a weighted sum of pure open string amplitudes with the graviton (or dilaton) replaced by a pair of collinear gauge bosons, each of them carrying exactly one-half of its momentum, {\it cf.\/} Eq. \split.

In order to express the partial amplitude $A(1,2,\dots,N{-}2;q)$ in terms of $N$-point open string amplitudes, we need to analyze the phase factors.
For a given $x_l<\xi<x_{l+1}$ with $l=2,\ldots,N-3$ the phase factor \PHASE\ in the integrand  can be accommodated by considering respective contours in the complex  $\eta$--plane.
After fixing the position of the first open string vertex at $x_1=-\infty$ we have the situation depicted in Fig. 1.

\medskip
\centerline{\epsfxsize=0.8\hsize\epsfbox{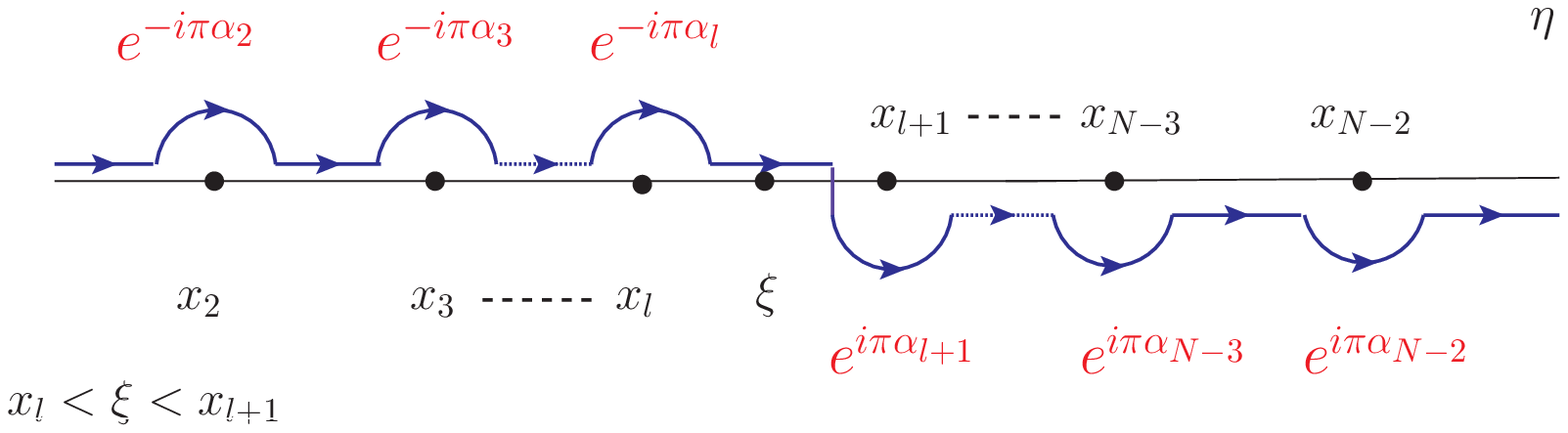}}
\noindent{\ninepoint\sl \baselineskip=8pt \centerline{{\bf Figure 1}: \sl
Complex $\eta$--plane and contour integrations. Here $\alpha_l\equiv\alpha'p_lq=2\alpha'k_lk_N$.}}

\bigskip\bigskip
\noindent
Quite generally, around all open string vertex positions $x_l<\xi$  the contour goes clockwise, while for $x_l>\xi$ anti--clockwise.
In either case we can deform the contour to the left or right. To obtain a minimal set of integration regions for $x_2<\xi<x_{\ceiling{\fc{N}{2}}}$ we move the contours to the left, {\it cf.\/} Fig. 2.

\bigskip
\centerline{\epsfxsize=0.8\hsize\epsfbox{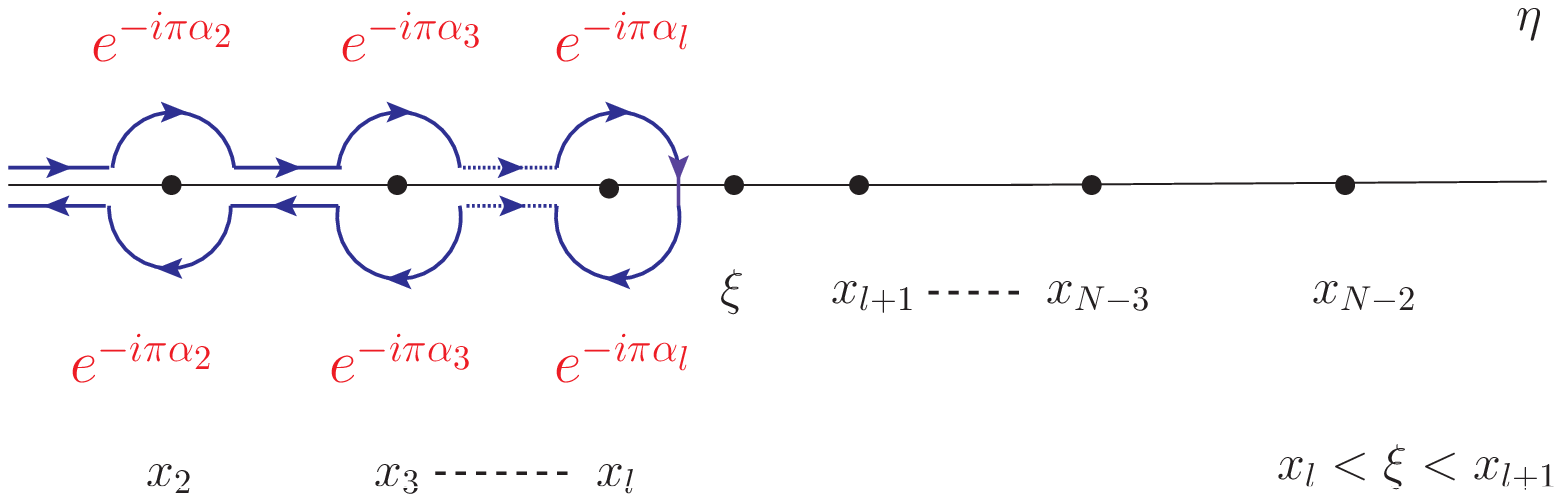}}
\noindent{\ninepoint\sl \baselineskip=8pt \centerline{{\bf Figure 2}: \sl
Contour deformation in complex $\eta$--plane.}}

\bigskip\bigskip
\noindent
On the other hand, for $x_{\ceiling{\fc{N}{2}}}<\xi<x_{N-2}$ we swap the contour to the right.
This way for each region $x_l<\xi<x_{l+1}$ with $l=2,\ldots,\ceiling{\fc{N}{2}}-1$
we obtain a residual contour of $l-1$ loops starting from $x_1=-\infty$ and encircling the $l-1$
points  $x_2,\ldots,x_l$. On the other hand, for each region $x_l<\xi<x_{l+1}$ with
$l=\ceiling{\fc{N}{2}},\ldots,N-3$ we get a contour of $N-2-l$ loops starting from $+\infty$ and encircling the $N-2-l$ points  $x_{N-2},\ldots,x_{l+1}$.
In total we obtain $(\ceiling{\fc{N}{2}}-2)(\floor{\fc{N}{2}}-1)$ terms:
\eqnn\FINAL{
$$\eqalignno{A(1&,2,\dots,N{-}2;q)=&\FINAL\cr
&=\sum_{l=2}^{\ceiling{\fc{N}{2}}-1}\sum_{i=2}^l
\sin\lf(\pi\sum_{j=i}^ls_{j,N-1}\ri)\ A(1,\ldots,i-1,N,i,\dots,l,N-1,l+1,\ldots,N-2)\cr
&+\sum_{l=\ceiling{\fc{N}{2}}}^{N-3}\sum_{i=l+1}^{N-2}
\sin\lf(\pi\!\sum_{j=l+1}^is_{j,N-1}\ri)\ A(1,\ldots,l,N-1,l+1,\ldots,i,N,i+1,\ldots,N-2) ,}
$$}
where $s_{i,j}\equiv s_{ij}= 2\alpha'k_ik_j$. On the r.h.s., according to \split\ $k_{N{-}1}=k_N=q/2$, and the helicities of  respective (labeled by $N{-}1$ and $N$, respectively) gauge bosons  are determined by the graviton ($--$ or $++$), or by the dilaton ($+-$ or $-+$). Note that in the zero slope $\ap\to 0$ limit
$\sin(\pi s_{kl})\to \pi s_{kl}$  all $N$--point open string amplitudes become pure Yang--Mills   subamplitudes:
\eqnn\FINALL{
$$\eqalignno{~~A_{\rm EYM}&(1,2,\dots,N{-}2;q)=&\FINALL\cr
=\pi\!\!&\sum_{l=2}^{\ceiling{\fc{N}{2}}-1}\sum_{i=2}^l
\lf(\sum_{j=i}^ls_{j,N-1}\ri)\ A_{\rm YM}(1,\ldots,i-1,N,i,\dots,l,N-1,l+1,\ldots,N-2)\cr
+\,\pi\!\!&\sum_{l=\ceiling{\fc{N}{2}}}^{N-3}\sum_{i=l+1}^{N-2}
\lf(\sum_{j=l+1}^is_{j,N-1}\ri)\ A_{\rm YM}(1,\ldots,l,N-1,l+1,\ldots,i,N,i+1,\ldots,N-2) .}
$$}
 
Let us consider some examples with a small number of external particles\foot{A formula similar to Eq. \FINALL\ has been considered before in Ref. \ChenCT.}. For $N=5,6$ and $N=7$ our formula \FINAL\ yields:
\eqn\three{\hskip-5.6cm A(1,2,3;q)=\sin(\pi s_{24})\ A(1,5,2,4,3)\ ,}
\eqn\four{\hskip-1.3cm A(1,2,3,4;q)=\sin(\pi  s_{25})\ A(1,6,2,5,3,4)+\sin(\pi s_{45})\ A(1,2,3,5,4,6)\ ,}
\eqn\five{\eqalign{
A(1,2,3,4,5;q)&=\sin(\pi s_{26})\ A(1,7,2,6,3,4,5)+\sin(\pi s_{36})\ A(1,2,7,3,6,4,5)\cr
&+\sin[\pi(s_{36}+s_{26})]\ A(1,7,2,3,6,4,5)+\sin(\pi s_{56})\ A(1,2,3,4,6,5,7)\ .}}
The first two cases have already been worked out in \StiebergerHQ. However, let us investigate their
structure in more detail.

In order to make connection with EYM theory, let us take the zero slope limit of
Eq. \three, for the same helicity configuration as in Eq. \example:
\eqn\eyex{
A(1^+,2^+,3^-;q^{--})=\pi\ s_{24}\ A_{\rm YM}(1^+,5^-,2^+,4^-,3^-)\qquad (\alpha'\to 0)\ .}
The Yang-Mills amplitude is the maximally helicity violating
\eqn\eyey{A_{\rm YM}(1^+,5^-,2^+,4^-,3^-)= 4\ {[12]^4\over [1q][q3][13][2q]^2}}
where we set $|4]=|5]=\fc{|q]}{\sqrt{2}}$, {\it cf.} Eq. \split. After using $s_{24}=\fc{s_{2q}}{2}\equiv\fc{t}{2}$ and momentum conservation, we find that the graviton amplitude agrees with Eq. \example, up to an overall factor which is necessary in order to convert string mass units into the gravitational
$\sqrt{\kappa}$. On the other hand, at the full--fledged string level of Eq. \three, we can use the expression for the five--point  open superstring amplitude $A(1,5,2,4,3)$ \MafraNVi, and take its collinear limit, \ie $s_{12}=s,\;s_{23}=u,\;s_{34}=\fc{s}{2},\;
s_{45}=0$ and $s_{51}=\fc{u}{2}$, {\it cf.} Eq. \split, to obtain
\eqn\Mixed{
A(1,2,3;q)=\pi\ \fc{t}{2}\ A_{\rm YM}(1,5,2,4,3)\ \sv\lf\{F\lf(\fc{s}{2},\fc{u}{2}\ri)\ri\}\  \ ,}
where\foot{According to the definition in Eq. \AMPLITUDE\ we have $F_5=\fc{\pi t}{2}\ \sv\lf\{F\lf(\fc{s}{2},\fc{u}{2}\ri)\ri\}$.} 
\eqn\formv{
F(s,u)=\fc{\Gamma(1+s)\ \Gamma(1+u)}{\Gamma(1+s+u)}}
is the four--point open superstring formfactor and sv is the single--valued projection\foot{It is worth mentioning that $\sv\{F(s,u)\}=\sv\{F(s,t)\}=\sv\{(F(t,u)\}$.}, previously discussed in the string context in \refs{\StiebergerWEA,\StiebergerHBA}. Alternatively, we can use the well--known relation \BernQJ
\eqn\bcjj{s_{25}\ A_{\rm YM}(1,5,2,4,3)=-s_{12}\ A_{\rm YM}(1,2,3,4,5)-(s_{12}+s_{23})\ A_{\rm YM}(1,3,2,4,5),}
to rewrite \Mixed\ as:
\eqn\Mixedd{
A(1,2,3;q)=-\pi\ \lf[\ s\ A_{\rm YM}(1,2,3,4,5)-t\ A_{\rm YM}(1,3,2,4,5)\ \ri]\
\sv\lf\{F\lf(\fc{s}{2},\fc{u}{2}\ri)\ri\}\ .}
Note that in \Mixed\ and \Mixedd\ the single--valued projection eliminates all powers
of $\zeta_2$ in the $\ap$--expansion of the amplitude $A(1,2,3;q)$. This is special for final states with  two or three gauge bosons; with more gauge bosons in the final state, the amplitudes will start receiving contributions from the $\zeta_2(F_{\mu\nu})^4$ effective interaction terms.

Next, let us discuss the five--point amplitude \four.
Here, we use the expressions for  six--point open superstring amplitudes $A(1,2,3,5,4,6)$
and $A(1,6,2,5,3,4)$ \MafraNVi, and take their collinear limit. Six--point string functions depend on nine kinematic invariants: $s_{i,i+1}=\ap(k_i+k_{i+1})^2$, $i=1,\ldots, 6\ \mod\ 6$, and $t_1=\ap(k_1+k_2+k_3)^2,\ t_2=\ap(k_2+k_3+k_4)^2,\ t_3=\ap(k_3+k_4+k_5)^2$. In the collinear limit of Eq. \split,
$s_{12}=s_1,\ s_{23}=s_2,\;s_{34}=s_3,\;s_{45}=\fc{s_4}{2},\;
s_{56}=0,\;  s_{61}=\fc{s_5}{2}$ and $t_1=s_4,\;t_2=s_5,\;
 t_3=\fc{s_1}{2}+\fc{s_3}{2}$ (\cf \StiebergerTE), where $s_i\equiv s_{i,i+1}$, $i=1,\ldots, 5\ \mod\ 5$, are the five-point kinematic invariants.  In this way, we obtain
\eqnn\MIXED{
$$\eqalignno{A(1,2,3,4;q)&=\pi \lf\{\ F_{6a}\ A_{\rm YM}(1,6,2,5,3,4)\ri.&\MIXED \cr
&\lf.+F_{6b}\ \lf[\ A_{\rm YM}(1,6,5,2,3,4)+A_{\rm YM}(1,5,6,2,3,4)\ \ri]\ \ri\}+(1\leftrightarrow 3)\ ,}$$ }
with the $\ap$ expansions:
\eqnn\Funcs{
$$\eqalignno{
F_{6a}&=-\h\ \fc{s_5\ (s_1-s_3-s_4)}{s_4}\ \lf\{\ 1-\fc{\zeta_2}{2}\ \lf(\ 2\ s_1s_2-s_1s_4+s_3s_4-s_4s_5\ \ri)\ \ri\}+\Oc(\ap^4)\ ,\cr
F_{6b}&=\h\ \fc{s_5}{s_4}\ \lf\{\ (s_4+s_5)-\zeta_2\ \lf(\ s_2s_3s_4+s_1s_2s_5-s_2s_4s_5\ \ri)\ \ri\}+\Oc(\ap^4)\ .&\Funcs}$$}

The collinear limits of Yang--Mills amplitudes have been studied for a long time \notation. Partial amplitudes with adjacent (in the gauge group trace factor) gauge bosons, like number 4 and 5 on the r.h.s.\ of Eqs. \Mixedd\ and \bcjj, contain collinear divergences and, at the leading order, factorize into a divergent factor times the amplitude with the collinear pair replaced by a single particle \notation. These leading divergences cancel in Eq. \Mixedd, as it is clear from Eq. \bcjj. The collinear limits on the r.h.s.\ of Eq. \FINAL\ do not contain singularities because the relevant gauge bosons are not adjacent. It would be very useful to have some compact formulas for such limits. They would require understanding the case of adjacent collinear gauge bosons at the subleading level\foot{In the soft limit, subleading divergences of graviton and Yang-Mills amplitudes have been recently studied in Refs. \CachazoFWA\ and \CasaliXPA, respectively.}.
{}For full--fledged string amplitudes, one  also needs collinear limits of string formfactors, as in Eq. \Funcs, to all orders in $\ap$.

It is tempting to think of the two gauge bosons -- that substitute for the  graviton or dilaton in the scattering amplitudes -- as their constituent particles. The idea that gravity may be induced by some other interactions was contemplated long ago by Andrei Sakharov \sakh\ (see also \Veneziano), but it has never been implemented in a satisfactory theoretical framework.
It is clear that Weinberg--Witten theorem \WeinbergKQ\ represents a significant (but hopefully surmountable) obstacle to graviton compositeness, so it would be interesting to see how it works in the context of amplitude relations derived in this work.
In order to seriously consider gravitons as bound states of gauge bosons, one would have to understand the monodromy factors of
Eq. \FINAL\ in terms of two-particle wave functions of the underlying gauge (open superstring) theory.

\vskip0.5cm
\goodbreak
\leftline{\noindent{\bf Acknowledgments}}

\noindent
We are grateful to CERN Theory Unit,  where a substantial portion this work was performed, for its hospitality and financial support.
This material is based in part upon work supported by the National Science Foundation under Grants No.\ PHY-0757959 and PHY-1314774.   Any
opinions, findings, and conclusions or recommendations expressed in
this material are those of the authors and do not necessarily reflect
the views of the National Science Foundation.

\listrefs

\end